# Detector Backgrounds at the Higgs Factory Muon Collider: MARS *vs* FLUKA

**N.V. Mokhov, S.I. Striganov[a], I.S. Tropin**

*Fermi National Accelerator Laboratory,*
*Batavia, IL 60510-5011, USA*
[a]*E-mail:* `striganov@fnal.gov`

**T.W. Markiewicz, T. Maruyama**

*SLAC National Accelerator Laboratory,*
*Menlo Park, CA 94025, USA*

ABSTRACT: Simulations for the 125-GeV Higgs Factory (HF) Muon Collider (MC) have shown large background particle loads on the collider detector. To verify level, source and composition of background calculations were performed using FLUKA and MARS codes for two shielding configurations. After comprehensive tuning of muon beam parameters, geometry setups and scoring procedures, background particle distributions at the detector entrance were simulated and compared. The spatial distributions and energy spectra of background particles obtained by two codes are rather similar. Average numbers of background particles simulated using MARS and FLUKA agree within a factor of two.

# Introduction

A Higgs Factory (HF) Muon Collider (MC) offers unique possibilities for studying the Higgs boson [1]. The radiation background produced by muon decays is the fundamental and critical issue in determining the feasibility of HF and its detectors. Muon decays are the major source of the detector background at muon colliders [2-4]. The decay length for a 62.5 GeV muon is $3.9 \cdot 10^5$ m. With $2 \cdot 10^{12}$ muons per bunch, this results in $10^7$ decays per meter in a single pass. The HF ring considered here is designed [5] for 1000 to 2000 turns per a store with 30 stores per second. Electromagnetic/hadronic showers induced by decay electrons in the collider components create difficulties with reconstruction of tracks, deteriorates detector resolution and produces radiation damage in detector components. Reliable calculations of background loads in HF detector are important for evaluation of a feasibility of this project. In this paper, FLUKA [6] and MARS15 [7,8] code inter-comparison has been performed with respect to the HF detector background predictions.

# Setting the simulations

First, a thorough verification of the HF MC lattice [9] and superconducting magnet protective components [10] used by FLUKA and MARS15 were performed. Then, we have made identical in both the codes the muon beam parameters and descriptions (positions, geometry, materials and magnetic fields) of the magnets and machine-detector interface (MDI). After thorough test runs, the scoring procedures were tuned up to calculate the same background particle characteristics.

One of the crucial background reduction components is a nozzle sitting on the beam very close to the Interaction Point (IP) [2-4, 10]. Simulations were performed for its two configurations, v2 and v7x2s4, which differ near IP at r<5cm in the ±100cm region longitudinally as shown in Fig. 1. Both setups have the same outer surface configuration: a 15-degree slope for the first 120 cm from IP followed by a 5-degree slope up to 350 cm. Other details of the nozzle and MDI overall are given in Ref. [4, 10].

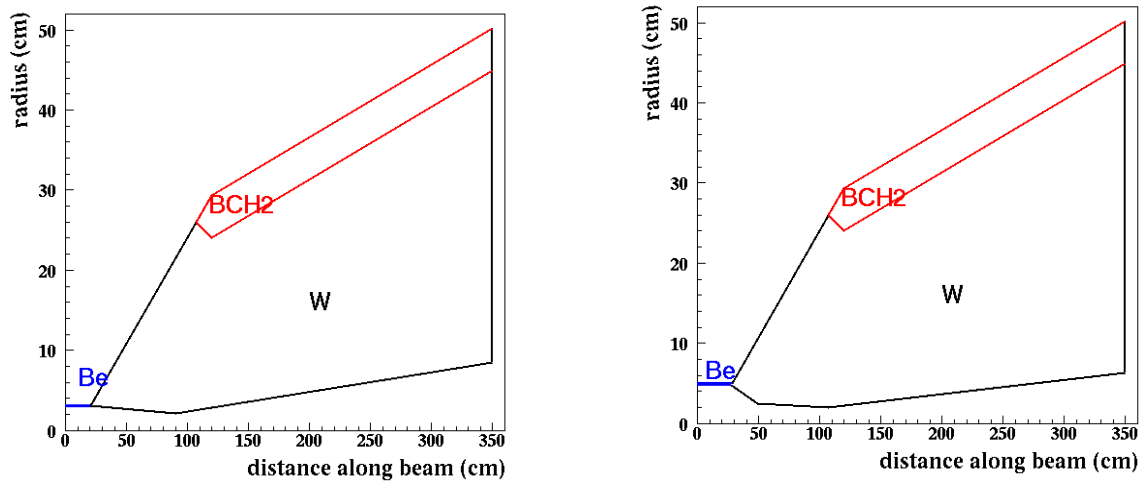

**Figure 1.** Nozzle configurations v2 (left) and v7x2s4 (right).

Both FLUKA and MARS15 are capable to model all particle physics and transport down to $10^{-5}$ - $10^{-3}$ eV for neutrons and 1 keV for all other particles. Calculated with MARS15 with these default low-energy thresholds photon and neutron fluence isocontours in the central detector region for v7x2s4 are presented in Figs. 2 and 3.

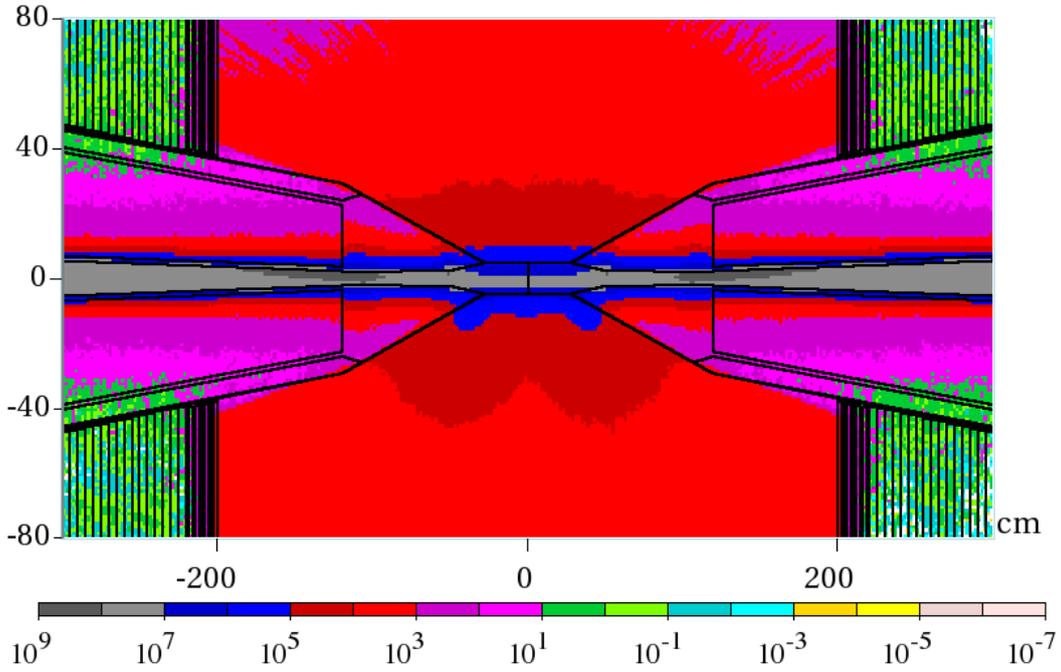

**Figure 2.** MARS15-calculated photon fluence (1/cm$^2$/BX).

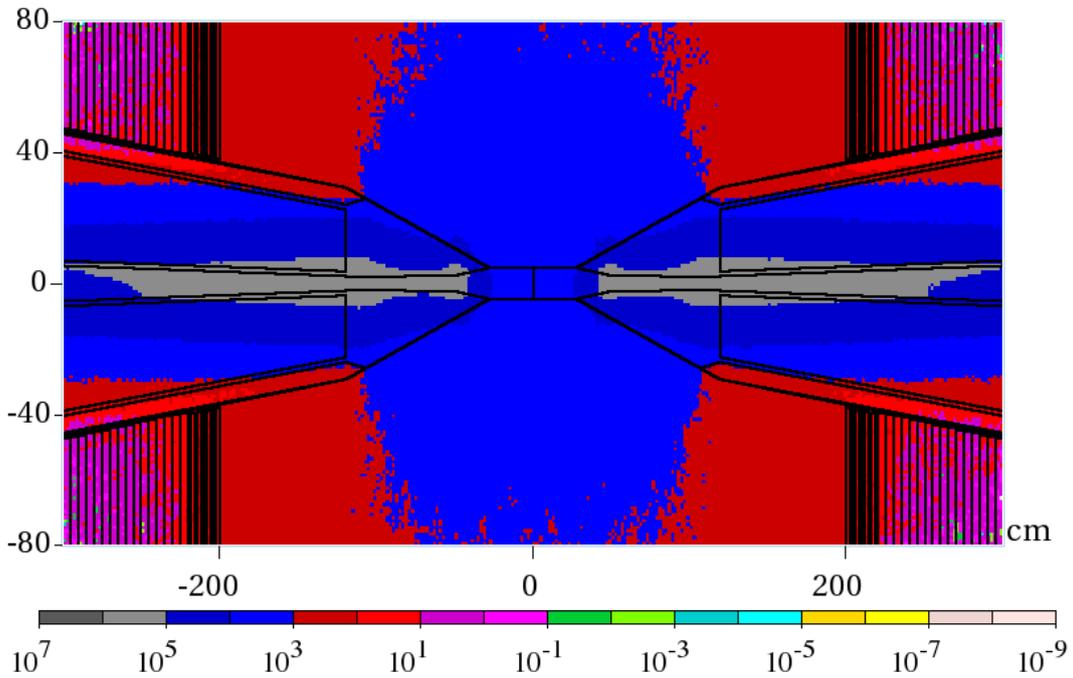

**Figure 3.** MARS15-calculated neutron fluence (1/cm$^2$/BX).

## Results of inter-comparison

To the purpose of the code inter-comparison, the cutoff energies in both the codes have been substantially increased to 0. 1 MeV for photons and neutrons and to 0. 5 MeV for electrons which allowed to drastically reduce the calculation time.

FLUKA and MARS calculated numbers of background particles entering the detector are presented in Tables 1 and 2 for configurations v2 and v7x2s4, respectively. It is seen that both the codes predict similar numbers for photons, electrons, neutrons and charged hadrons. Photon and electron numbers are about ten times larger than those calculated for the 1.5-TeV MC [3]. For the v2 configuration, about 70% of background particles are coming into detector near the first quadrupole. About $2.2 \times 10^5$ electrons from muon decay (per bunch crossing) come into the detector through Be beam pipe near IP.

Table 1: Number of particles entering detector per bunch crossing for v2 nozzle.

| Particle | Threshold (MeV) | FLUKA | MARS |
|---|---|---|---|
| photons | 0.1 | $1.8\ 10^9$ | $1.6\ 10^9$ |
| e± | 0.5 | $7.0\ 10^7$ | $6.1\ 10^7$ |
| neutrons | 0.1 | $1.0\ 10^8$ | $8.6\ 10^7$ |
| charged hadrons | 1.0 | $7.0\ 10^4$ | $5.1\ 10^4$ |

Table 2: Number of particles entering detector per bunch crossing for v7x2s4 nozzle.

| Particle | Threshold (MeV) | FLUKA | MARS |
|---|---|---|---|
| photons | 0.1 | $6.3\ 10^8$ | $3.0\ 10^7$ |
| e± | 0.5 | $2.1\ 10^7$ | $0.8\ 10^7$ |
| neutrons | 0.1 | $6.7\ 10^7$ | $4.2\ 10^7$ |
| charged hadrons | 1.0 | $3.1\ 10^4$ | $1.1\ 10^4$ |

The configuration v7x2s4 also includes additional shielding around the first quadrupole. The tungsten mask radii were decreased from $5\sigma_{x,y}$ in the v2 configuration to $4\sigma_{x,y}$. The beam pipe radius near IP was enlarged from 3 to 5 cm. The inner nozzle surface was reconfigured to prevent direct hits of the beam pipe by decay electrons (Fig.1). One can see from the Tables that with this configuration improvement, both the codes predict similar substantial reduction of the background particle load on the detector.

The background particle distributions of the detector entrance points are shown in Fig.2. For clarity, only background produced by $\mu^-$ decay from the right side is shown. To make the comparison clearer, MARS results were renormalized to have the same average particle numbers as FLUKA backgrounds (divided by 2.1 for photons, 2.6 for electrons/positrons, and 1.6 for neutrons). It is seen that the distribution shapes are similar for background particles produced downstream of the IP, with more particles coming to the detector from the IP upstream for the FLUKA simulations. Note, that it is not practically important, because much more particles are produced in this region by $\mu^+$ decay from the left side.

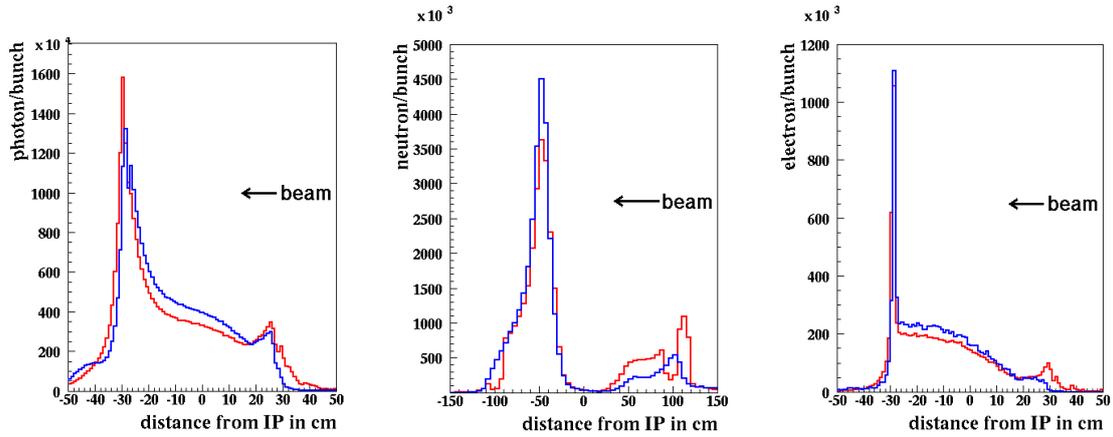

Fig.2. Entrance to the detector of photons (left), neutrons (central) and electrons (right). FLUKA – red, MARS -blue.

Energy spectra of the background particles are presented in Fig.3. Again, MARS background was renormalized to the same number of particles as FLUKA. The spectrum shapes are rather similar, with a little bit more energetic tails obtained with FLUKA.

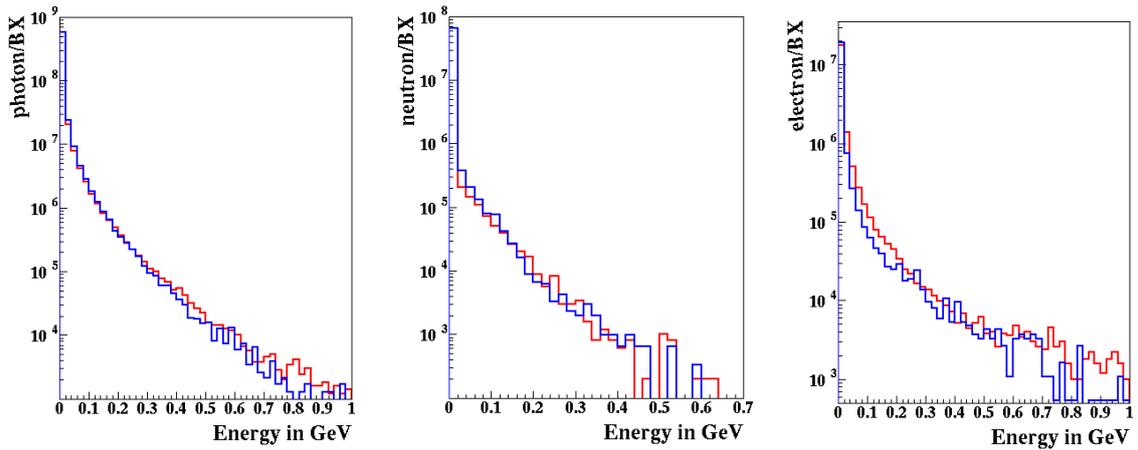

Fig.3. Energy spectra of particles entering the detector: photon(left), neutron(central), electron (right). FLUKA – red, MARS -blue.

## Analysis

Overall, shown in the tables and figures results predicted by both the codes are very close to each other for the simpler v2 configuration. As for the more efficient configuration v7x2s4, MARS predicts 1.5 to 3 times lower background load. We attribute this to a better efficiency of the tighter magnet protection system (masks and liners) – described in the previous paragraph - individually fine-tuned for each the magnet in the HF lattice in MARS, not yet implemented in the FLUKA model.

On the physics side, most of the background photons are produced in showers generated by decay electrons near the 107-cm nozzle tip before IP and -49cm tip after IP and entered detector near the beryllium beam pipe end (-27 cm). These photons are produced at large angles. Simulation of such photon production and near-surface transport in a strong magnetic field require a special attention and should be verified in the codes. For production, MARS uses an approximation of the Tsai formula [11], the EGS mode in MARS uses approach [12], and FLUKA manual just states "the angular distribution of bremsstrahlung photons is sampled accurately".

The hit distributions in the detector could not be compared in this study because of too many differences for that in the approaches used in both the codes. It is important that in both simulations, most of the hits are generated by photons. Therefore, it follows from the results described in this paper that the uncertainty in the hit predictions in the HF detector should not exceed a factor of two.

## Acknowledgments

This document was prepared using the resources of the Fermi National Accelerator Laboratory (Fermilab), a U.S. Department of Energy, Office of Science, HEP User Facility. Fermilab is managed by Fermi Research Alliance, LLC (FRA), acting under Contract No. DE-AC02-07CH11359.